\documentclass[aps,prd,twocolumn,superscriptaddress,nofootinbib,floatfix]{revtex4-2}
\usepackage{amsmath,amssymb,bm}
\usepackage{array}
\usepackage{graphicx}
\usepackage{microtype}
\usepackage{xcolor}
\usepackage[colorlinks=true,linkcolor=blue!55!black,citecolor=blue!55!black,urlcolor=blue!55!black]{hyperref}
\usepackage[nameinlink,capitalise]{cleveref}
\crefname{figure}{Fig.}{Figs.}
\crefname{section}{Sec.}{Secs.}
\crefname{equation}{Eq.}{Eqs.}
\crefname{table}{Table}{Tables}
\Crefname{figure}{Fig.}{Figs.}
\Crefname{section}{Sec.}{Secs.}
\Crefname{equation}{Eq.}{Eqs.}
\Crefname{table}{Table}{Tables}
\newcommand{\OXg}{\mathcal O_{X\gamma4}}
\newcommand{\OT}{\mathcal O_T}
\newcommand{\Osev}{\mathcal O_{17}}
\newcommand{\Otwo}{\mathcal O_{20}}
\newcommand{\Sp}{\Sigma'}
\newcommand{\DD}{\Delta}
\begin{document}

\title{Elastic toroidal vector dark matter through a dark photon in the LUX-ZEPLIN high recoil window}

\author{Imtiaz Khan}
\email{ikhanphys1993@gmail.com }
\affiliation{Department of Physics, Zhejiang Normal University, Jinhua, Zhejiang 321004, China}
\affiliation{Research Center of Astrophysics and Cosmology, Khazar University, Baku, AZ1096, 41 Mehseti Street, Azerbaijan}

\author{Salvatore Capozziello}
\email{capozziello@na.infn.it}
\affiliation{Dipartimento di Fisica ``E. Pancini", Universit\`a di Napoli ``Federico II", Complesso Universitario di Monte Sant’ Angelo, Edificio G, Via Cinthia, I-80126, Napoli, Italy,}
\affiliation{Istituto Nazionale di Fisica Nucleare (INFN), sez. di Napoli, Via Cinthia 9, I-80126 Napoli, Italy,}
\affiliation{Scuola Superiore Meridionale, Via Mezzocannone 4, I-80134, Napoli, Italy.}

\author{G. Mustafa}
\email{gmustafa3828@gmail.com}
\affiliation{Department of Physics, Zhejiang Normal University, Jinhua, Zhejiang 321004, China}

\author{Farkhod Botirov}
\email{f.botirov@nuu.uz}
\affiliation{National University of Uzbekistan, Tashkent 100174, Uzbekistan}

\author{Ahmadjon~Abdujabbarov}
\email{ahmadjonab@gmail.com}
\affiliation{School of Physics, Harbin Institute of Technology, Harbin 150001, People's Republic of China}

\author{Farruh~Atamurotov}
\email{atamurotov@yahoo.com}
\affiliation{Kimyo International University in Tashkent, Shota Rustaveli str. 156, Tashkent 100121, Uzbekistan}

\author{Phongpichit Channuie}
\email{phongpichit.ch@mail.wu.ac.th (Corresponding Author)}
\affiliation{School of Science \& College of Graduate Studies, Walailak University, Nakhon Si Thammarat, 80160, Thailand}

\begin{abstract}
The 2026 LUX-ZEPLIN search reports one nuclear recoil candidate at $248\pm23_{\rm stat}\pm23_{\rm sys}~\mathrm{keV}$ in an extended window reaching about $270~\mathrm{keV}$. We consider elastic scattering of neutral complex vector dark matter through its dimension-six toroidal electromagnetic moment and find appreciable spectral weight in this recoil region. The nonrelativistic reduction correlates $\mathcal O_{17}$ and $\mathcal O_{20}$, relating a coherent charge branch to a transverse spin tensor branch. For $m_X=1~\mathrm{TeV}$, natural xenon exhibits a broad high recoil maximum at $204.8$ to $206.6~\mathrm{keV}$ and retains $80.4$ to $85.4\%$ of this maximum at $248~\mathrm{keV}$. Resolving the interaction through a dark photon relates the high-to-low recoil ratio to the mediator mass. A renormalizable $SU(2)_X\times U(1)_D$ realization stabilizes the complex vector by a residual $Z_3$ gauge symmetry and generates the toroidal current through vectorlike messengers carrying a physical CP phase. For a benchmark with $m_X=0.998~\mathrm{TeV}$ and $m_{A'}=0.450~\mathrm{GeV}$, $R(248)/R(50)=1.43$ for GCN and $1.21$ for JJ55, while the odd $A$ xenon fraction remains above $99.9\%$. The response level ratio between the $350$ to $590~\mathrm{keV}$ and $100$ to $270~\mathrm{keV}$ intervals is about $0.46$ at this benchmark. The tensor branch is absent for spin zero argon, whereas annual modulation remains at the few percent level. The resulting spectral, isotope, target, timing, mediator, and high energy dependences give correlated tests of an elastic tensor interpretation of the LZ recoil.
\end{abstract}
\maketitle

\section{Introduction}
\label{sec:intro}

The latest LUX-ZEPLIN analysis extends the nuclear recoil search to about $270~\mathrm{keV}$ using $2.84~\mathrm{tonne\,yr}$ and reports one candidate at $248\pm23_{\rm stat}\pm23_{\rm sys}~\mathrm{keV}$. The background only hypothesis has a global tension of $2.6\sigma$, while the largest local significance among the tested interactions is $3.4\sigma$~\cite{LZ2026highE}. Earlier LZ analyses established sensitivity to nonrelativistic interactions over an enlarged recoil range~\cite{LZ2024nreft,LZ2024covariant}. LUX and XENON obtained related results from effective interaction and inelastic analyses~\cite{LUX2021eft,XENON2024eft}, and broader studies emphasized the relevance of an extended recoil window for momentum dependent interactions~\cite{Bozorgnia2018dvr}. The statistical evidence remains limited, but the enlarged window poses a definite particle physics question. Conventional elastic coherent scattering is concentrated at lower recoil energy. An interaction with appreciable support near $250~\mathrm{keV}$ must instead rely on kinematics, momentum dependence, nuclear structure, or a combination of these effects. We therefore focus on this extended recoil regime and formulate response-level tests that can be applied to future high-recoil data.

The LZ analysis provides a direct example of this situation. Its high energy search probes momentum dependent covariant interactions and uses $\mathcal L_{10}^{s}$ as an elastic case with a broad double peaked recoil spectrum~\cite{LZ2026highE}. The conventional NREFT basis used in these analyses is constructed from $\bm q$, $\bm v^\perp$, the dark matter spin vector, and the nucleon spin. For spin $1$, the polarization space contains in addition an independent rank two tensor, enlarging the Galilean operator basis~\cite{Catena2019spin1,Gondolo2021spin}. The high recoil event therefore provides motivation to examine a response sector specific to vector dark matter and not contained in the usual spin vector operators.

Most current interpretations shift the recoil spectrum through dark sector or incident particle kinematics. Endothermic scattering introduces a positive internal splitting, as developed in the inelastic dark matter framework~\cite{TuckerSmith2001hy,TuckerSmith2004jv,Barello2014eaa}. High recoil searches are especially sensitive to this kinematic threshold~\cite{Bramante2016rdh,Chang2009idm}, and the present LZ event has been considered in this setting~\cite{Su2026lz}. A broader phenomenological study likewise finds that ordinary elastic coherent scattering is weighted toward lower recoil, whereas momentum or spin dependent interactions can retain support at higher energies~\cite{DiMauro2026lz}. Higgsino interpretations relate the high recoil region to a splitting within an electroweak neutral sector~\cite{Freese2026lz,FanReece2026lz,WuZhangZhu2026lz}, with related electroweak studies constraining the corresponding neutral-state spectrum~\cite{DuWang2026lz,Smirnov2026lz}. Pseudo Dirac and singlet doublet constructions realize the threshold in explicit gauge or fermion sectors~\cite{Borah2026,DuHuangXie2026}, while collider and extended neutralino analyses provide complementary particle physics information~\cite{CheungKangKumar2026,BisalCaoLi2026}. Dark photon models can also place the recoil close to a kinematic endpoint~\cite{Yamashita2026lz,Zhu2026lightDP}. Exothermic scattering releases internal energy~\cite{BaerBarger2026,deLima2026}, whereas boosted populations replace the high speed halo tail with a nonthermal incident flux~\cite{Alhazmi2026,HeikinheimoZimmermann2026}. Absorption and atmospheric neutrino upscattering provide additional possibilities~\cite{Lou2026lz,Jeesun2026nu}. Elastic pseudoscalar scattering and an elastic isovector interaction illustrate separately that a high recoil feature need not require excitation of a dark state~\cite{Unwin2026lz,Khan2026L6v}. Elastic spin scattering, transition dipoles, geometric inelastic models, and sideband analyses further distinguish these possibilities~\cite{ElahiSchwaller2026,He2026transition,AhmedLeontaris2026,Rodd2026sideband,DentNewstead2026}. The relevant question is therefore not only whether one recoil energy can be reproduced. A consistent mechanism also relates the low and high recoil populations to isotope content, target spin, mediator propagation, timing, and the response above the nominal analysis window. These observables depend on different parts of the scattering dynamics and help separate kinematic thresholds from nuclear structure and momentum dependent interactions.

We consider a neutral vector particle whose spin tensor response contributes in the LZ high recoil region and derive the associated observables. A spin $1$ state possesses a rank two traceless polarization tensor absent for spin $0$ and spin $1/2$ dark matter~\cite{Catena2019spin1,Gondolo2021spin,Liang2025vector}. The electromagnetic form factors of a neutral complex vector include a dimension six toroidal moment~\cite{Hisano2020vector,Chu2023multipole,Bertuzzo2024complex}. Its nonrelativistic reduction generates the correlated operators $\Osev$ and $\Otwo$~\cite{Liang2025vector}. Their electromagnetic origin fixes the relation between the low recoil coherent contribution and the high recoil tensor spin contribution.

We formulate this interaction for the LZ high recoil candidate and embed it in a dark gauge sector. In xenon the spectrum contains a broad tensor maximum near $205~\mathrm{keV}$ and remains sizable at the candidate energy. The coherent-to-tensor crossover occurs near $60$ to $64~\mathrm{keV}$ and depends weakly on the mediator mass. The high to low recoil ratio instead depends on the mediator and yields a lower dark photon mass that can be inverted in the propagator transition region. The high recoil branch is carried mainly by $^{129}\mathrm{Xe}$ and $^{131}\mathrm{Xe}$, whereas a spin zero argon target lacks this branch. Annual modulation remains at the few percent level for elastic scattering. The response above $350~\mathrm{keV}$ provides another mediator-sensitive quantity. The gauge realization relates these observables to vector stability, the mediator mass, a physical CP phase, and a loop generated toroidal current.

We employ the public GCN and JJ55 xenon one body density matrices distributed with \texttt{dmscatter}~\cite{Gorton2022cpc,AbdelKhaleq2022natXe,AbdelKhaleq2024nuc}. GCN and JJ55 correspond to two shell model interactions and model spaces used in evaluating the xenon response functions, and their spread is used below as a measure of nuclear interaction dependence. For $^{40}\mathrm{Ar}$ we use SDPF-U and SDPF-MU. The Galilean operator basis follows the standard nonrelativistic effective theory construction~\cite{Fan2010gt,Fitzpatrick2012ix,Fitzpatrick2012ib}. Nuclear response functions and Lorentz to Galilean matching are taken from Refs.~\cite{Anand2013yka,DelNobile2018dfm,Gresham2014vja}. Interference and global response analyses provide complementary information on operator combinations~\cite{Catena2014uqa,Brod2017bsw}. Spin dependent xenon calculations give independent guidance for the odd isotopes~\cite{Menendez2012tm,Klos2013rwa,Vietze2014jga}. Chiral matching studies motivate retaining the momentum dependence of the nuclear current~\cite{Bishara2016hnh,Hoferichter2015chiral,Hoferichter2016analysis}, while coherent structure factor calculations provide additional finite momentum checks of the charge response~\cite{Hoferichter2018acd,AbdelKhaleq2025cevns}.

\section{Toroidal vector interaction}
\label{sec:operator}

\subsection{Electromagnetic moment and nonrelativistic matching}

Let $X_\mu$ denote a neutral complex vector field of mass $m_X$. The nonrelativistic polarization states can be represented by the symmetric traceless tensor
\begin{equation}
 \widetilde S_X^{ij}=\frac12\left(S_X^iS_X^j+S_X^jS_X^i\right)-\frac23\delta^{ij},
 \label{eq:tensor}
\end{equation}
where $S_X^i$ are the spin $1$ generators. The tensor defined in \cref{eq:tensor} appears in the two Galilean operators~\cite{Catena2019spin1,Liang2025vector}
\begin{align}
 \Osev &= i\frac{\bm q}{m_N}\!\cdot\!\widetilde{\bm S}_X\!\cdot\!\bm v_N^\perp\,1_N,
 \label{eq:o17}\\
 \Otwo &=-\frac{\bm q}{m_N}\!\cdot\!\widetilde{\bm S}_X\!\cdot\!
 \left(\frac{\bm q}{m_N}\times\bm S_N\right).
 \label{eq:o20}
\end{align}
Here $\bm q$ denotes the momentum transfer to the nucleus, $m_N$ the nucleon mass, $\bm S_N$ the nucleon spin, and $\bm v_N^\perp$ the Galilean invariant transverse velocity. Because \cref{eq:o17,eq:o20} contain explicit powers of momentum, their nuclear response has to be retained at finite momentum transfer.

We consider the operator
\begin{equation}
 \OXg=\partial_\nu\!\left(X^{\mu\dagger}X^\nu+X^{\nu\dagger}X^\mu\right)
 \partial^\lambda F_{\mu\lambda}.
 \label{eq:oxg}
\end{equation}
For a neutral complex vector, the toroidal electromagnetic operator can be written as~\cite{Chu2023multipole}
\begin{equation}
 \OT=X_\mu^\dagger X_\nu
 \left(\partial^\mu\partial_\rho F^{\rho\nu}
 +\partial^\nu\partial_\rho F^{\rho\mu}\right).
 \label{eq:toroidal}
\end{equation}
Integration by parts maps \cref{eq:oxg} to \cref{eq:toroidal}, up to a total derivative and the field strength convention. The interaction is therefore the toroidal moment in the usual electromagnetic classification. This moment is odd under charge conjugation, even under parity, and odd under CP~\cite{Chu2023multipole}. Its CP property constrains the ultraviolet realization below.

For a nucleon $N$, the corresponding nonrelativistic reduction is~\cite{Liang2025vector}
\begin{equation}
 \OXg\longrightarrow 2 e m_N^2\left(2Q_N\Osev+g_N\Otwo\right).
 \label{eq:matching}
\end{equation}
We take $Q_p=1$, $Q_n=0$, $g_p=5.59$, and $g_n=-3.83$. Removing the common coefficient in \cref{eq:matching}, we define
\begin{equation}
 c_{17}^N=2Q_N,\qquad c_{20}^N=g_N.
 \label{eq:coeff}
\end{equation}
The electromagnetic interaction thus fixes the relative coefficients of the charge and magnetic tensor contributions.

\subsection{Response dictionary}

The recoil rate follows directly from the spin $1$ dark matter response functions. Introducing $x=q^2/m_N^2$, the four nonzero response coefficients generated by \cref{eq:coeff} are~\cite{Liang2025vector}
\begin{align}
 R_M^{NN'}&=\frac{x}{6}v_T^{\perp2}c_{17}^Nc_{17}^{N'},
 \label{eq:rm}\\
 R_\DD^{NN'}&=\frac{x^2}{6}c_{17}^Nc_{17}^{N'},
 \label{eq:rd}\\
 R_{\Sp}^{NN'}&=\frac{x^2}{24}c_{20}^Nc_{20}^{N'},
 \label{eq:rs}\\
 R_{\Sp\DD}^{NN'}&=-\frac{x^2}{12}c_{20}^Nc_{17}^{N'}.
 \label{eq:rsd}
\end{align}
The operator $\Osev$ therefore contributes to both the coherent $M$ response and the orbital $\DD$ response, while interference between $\Osev$ and $\Otwo$ generates $\Sp\DD$. The four response channels are consequently fixed by the correlated $\Osev$--$\Otwo$ interaction.

For an isotope with spin $J_A$, the proton and neutron coefficients are contracted with the nuclear functions $W_K^{NN'}$. We denote the resulting contracted responses by $\widehat W_K$. After integration over the velocity distribution, the recoil shape is
\begin{equation}
 \mathcal R_A(E_R)=
 \frac{x}{6}\widehat W_M I_\perp
 +\frac{x^2}{6}\widehat W_\DD\eta
 +\frac{x^2}{24}\widehat W_{\Sp}\eta
 -\frac{x^2}{12}\widehat W_{\Sp\DD}\eta.
 \label{eq:rate}
\end{equation}
The required velocity moments are
\begin{align}
 \eta(v_{\min})&=\int_{v>v_{\min}}\frac{f_{\rm lab}(\bm v)}{v}\,d^3v,
 \label{eq:eta}\\
 \xi(v_{\min})&=\int_{v>v_{\min}}v f_{\rm lab}(\bm v)\,d^3v,\\
 I_\perp&=\frac{\xi-v_{\min}^2\eta}{c^2}.
 \label{eq:iperp}
\end{align}
For elastic scattering the minimum incident speed is
\begin{equation}
 v_{\min}(E_R)=\sqrt{\frac{m_AE_R}{2\mu_{XA}^2}},\qquad
 \mu_{XA}=\frac{m_Xm_A}{m_X+m_A}.
 \label{eq:vmin}
\end{equation}
The moments in \cref{eq:eta,eq:iperp}, together with the elastic threshold in \cref{eq:vmin}, follow from $v_T^{\perp2}=v^2-q^2/(4\mu_{XA}^2)$ and retain the transverse velocity contribution over the mass range considered. We evaluate $\eta$ analytically for a truncated Maxwell distribution and check the result independently by direct numerical quadrature.

The public density matrices contain the valence-space contribution. We restore the inert closed shell in the coherent response. For both xenon interactions the reconstructed $q\to0$ amplitudes then satisfy $F_M^{pp}(0)=Z^2$, $F_M^{nn}(0)=N^2$, and $F_M^{pn}(0)=ZN$. This normalization is relevant because the low recoil branch is associated with proton charge, whereas the high recoil branch is governed by the magnetic tensor response of the odd $A$ isotopes.

\subsection{Resolved dark photon and renormalizable gauge realization}

We resolve the electromagnetic current interaction by introducing a massive dark photon and write
\begin{equation}
 \begin{aligned}
 \mathcal L_{A'}&=g_4^D A'_\mu J_T^\mu+\epsilon e A'_\mu J_{\rm EM}^\mu,\\
 J_T^\mu&=\partial_\nu\left(X^{\mu\dagger}X^\nu+X^{\nu\dagger}X^\mu\right),
 \end{aligned}
 \label{eq:mediatorlag}
\end{equation}
Here $m_{A'}$ is the dark photon mass, $g_4^D$ denotes the resolved toroidal current coupling, and $\epsilon$ is the kinetic mixing parameter. Kinetic mixing provides the standard renormalizable portal between Abelian gauge sectors~\cite{Holdom1986,Fabbrichesi2020darkphoton}. For spacelike exchange the recoil rate is multiplied by
\begin{equation}
 P(q,m_{A'})=\left(\frac{m_{A'}^2}{m_{A'}^2+q^2}\right)^2.
 \label{eq:prop}
\end{equation}
The heavy mediator limit approaches the toroidal contact interaction, whereas a light mediator reduces the explicit momentum enhancement.

For the ultraviolet realization we take the dark gauge group
\begin{equation}
 G_D=SU(2)_X\times U(1)_D.
 \label{eq:gaugegroup}
\end{equation}
An $SU(2)_X$ quadruplet $\Phi_4$ acquires a vacuum expectation value $v_4$ in its extremal weight components. This breaking pattern leaves a residual $Z_3$ gauge symmetry under which the complex vector
\begin{equation}
 X_\mu=\frac{X^1_\mu-iX^2_\mu}{\sqrt2}.
 \label{eq:complexvector}
\end{equation}
The symmetry breaking pattern and gauge boson mass relation follow the established quadruplet realization of vector dark matter~\cite{ChenNomura2015SU2,Foguel2026vector}. We identify $X_\mu$ with the lightest state carrying a nontrivial $Z_3$ charge. The masses of $X_\mu$ and the neutral gauge boson are
\begin{equation}
 m_X=\frac{\sqrt3}{2}g_Xv_4,
 \qquad
 m_{X_3}=\frac32 g_Xv_4=\sqrt3\,m_X.
 \label{eq:su2masses}
\end{equation}
A second scalar $S$, with unit $U(1)_D$ charge and vacuum expectation value $v_D$, gives
\begin{equation}
 m_{A'}=g_Dv_D.
 \label{eq:darkphotonmass}
\end{equation}
The vector $X_\mu$ is neutral under $U(1)_D$. Consequently, the renormalizable minimal interaction $A'X^\dagger X$, which would introduce a softer coherent contribution, is absent.

The toroidal current is generated through heavy vectorlike messengers. We introduce two doublet and triplet sectors
\begin{equation}
 D_a\sim({\bf2},Q_a),\qquad T_a\sim({\bf3},Q_a),
 \qquad Q_2=Q_1+1,
 \label{eq:messreps}
\end{equation}
under $SU(2)_X\times U(1)_D$. Since ${\bf2}\otimes{\bf4}={\bf3}\oplus{\bf5}$, a $D$-$\Phi_4$-$T$ contraction is allowed. With $Q_D(S)=1$, the gauge invariant messenger Lagrangian contains
\begin{equation}
\begin{aligned}
\mathcal L_{\rm mess}\supset{}&-\sum_{a=1}^2\Big[M_{Da}\bar D_aD_a+M_{Ta}\bar T_aT_a\\
&+y_{La}\bar D_{aL}\Phi_4T_{aR}+y_{Ra}\bar D_{aR}\Phi_4T_{aL}\Big]\\
&-\kappa_D\bar D_2 S D_1-\kappa_T\bar T_2 S T_1+{\rm h.c.}
\end{aligned}
\label{eq:messlag}
\end{equation}
Complex Yukawa couplings provide the physical phase required by the CP odd toroidal moment. A rephasing invariant combination for each sector is
\begin{equation}
 \mathcal J_{{\rm CP},a}=
 \frac{\operatorname{Im}\!\left(y_{La}y_{Ra}^{*}M_{Da}^{*}M_{Ta}^{*}\right)}
 {|M_{Da}M_{Ta}|}.
 \label{eq:jcp}
\end{equation}
This CP structure agrees with the general neutral three vector analysis, where the corresponding CP odd form factor requires a physical complex phase~\cite{HernandezJuarez2021CPV}. Heavy messenger sectors generate electromagnetic multipoles for neutral vector states and provide the loop framework used here~\cite{Hisano2020vector,Chu2023multipole}. Dark sector effective field theory supplies the operator organization~\cite{Liang2023dseft,Bertuzzo2024complex}.

The messenger sector selects the CP odd dimension six current interaction
\begin{equation}
 \mathcal L_{\rm eff}\supset\frac{c_T}{M_F^2}J_S^\mu J^T_\mu,
 \qquad
 J_S^\mu=iS^\dagger\!\overleftrightarrow D^\mu S.
 \label{eq:currentcurrent}
\end{equation}
After $U(1)_D$ breaking, $J_S^\mu$ contains the term $g_Dv_D^2A'^\mu$. It follows that
\begin{equation}
 g_4^D=c_T\frac{g_Dv_D^2}{M_F^2},
 \qquad
 C_T=-\frac{\epsilon g_4^D}{m_{A'}^2}
 =-\frac{\epsilon c_T}{g_DM_F^2},
 \label{eq:ctmatching}
\end{equation}
where $\mathcal L_{\rm tor}=C_T\OXg$. The one loop matching coefficient can be parameterized as
\begin{equation}
 c_T=\frac{g_X^2}{16\pi^2}\,\mathcal J_{\rm CP}\,
 \mathcal F(r_i,\theta_i).
 \label{eq:loopcoefficient}
\end{equation}
The finite function $\mathcal F$ depends on messenger mass ratios and mixing angles. Its explicit form follows from a one loop matching calculation for a specified messenger spectrum, while the recoil analysis below depends only on the resolved current structure and mediator propagator.

\begin{table}[!tbp]
\caption{Reference dark sector benchmark. The displayed loop normalization uses $\mathcal F=1$, while $\mathcal F$ remains a spectrum dependent matching function.}
\label{tab:uvbenchmark}
\centering
\scriptsize
\begin{tabular}{lc@{\qquad}lc}
\hline\hline
Parameter & Value & Derived quantity & Value \\
\hline
$g_X$ & $0.80$ & $m_X$ & $0.998~\mathrm{TeV}$ \\
$v_4$ & $1.44~\mathrm{TeV}$ & $m_{X_3}$ & $1.728~\mathrm{TeV}$ \\
$g_D$ & $1.50\times10^{-4}$ & $m_{A'}$ & $0.450~\mathrm{GeV}$ \\
$v_D$ & $3.00~\mathrm{TeV}$ & $\mathcal J_{\rm CP}$ & $0.866$ \\
$M_F$ & $3.00~\mathrm{TeV}$ & $c_T$ & $3.51\times10^{-3}$ \\
$\delta_{\rm CP}$ & $\pi/3$ & $g_4^D$ & $5.26\times10^{-7}$ \\
\hline\hline
\end{tabular}
\end{table}

The benchmark in \cref{tab:uvbenchmark} lies in the resolved mediator regime relevant for direct detection. It gives $H=R(248)/R(50)=1.43$ and $\mathcal B=2.30$ for GCN, compared with $H=1.21$ and $\mathcal B=1.92$ for JJ55. For both nuclear calculations the odd $A$ fraction at $248~\mathrm{keV}$ remains above $99.9\%$. The characteristic non Abelian annihilation scale is
\begin{equation}
 \frac{g_X^4}{64\pi m_X^2}=2.39\times10^{-26}~\mathrm{cm^3\,s^{-1}},
 \label{eq:annscale}
\end{equation}
placing the TeV benchmark near the conventional thermal scale, up to group, threshold, and final state factors.

The recoil shape also constrains a possible minimal current contribution. Let $R_T$ denote the toroidal rate and $R_s$ a soft coherent component. At $E_L=50~\mathrm{keV}$ we define $\zeta=R_s(E_L)/R_T(E_L)$. Then
\begin{equation}
 H_{\rm tot}=\frac{H_T+\zeta H_s}{1+\zeta},
 \label{eq:contam}
\end{equation}
where $H_i=R_i(248)/R_i(50)$. If $H_s<1$, a high recoil dominated spectrum requires
\begin{equation}
 \zeta<\frac{H_T-1}{1-H_s}.
 \label{eq:zetabound}
\end{equation}
For a coherent $\mathcal O_1$ shaped contribution, $H_s\simeq10^{-3}$. At $m_X=1~\mathrm{TeV}$, \cref{eq:zetabound} gives $\zeta<1.14$ for GCN and $\zeta<0.814$ for JJ55. Thus the hard branch constrains not only the mediator mass but also the allowed size of softer ultraviolet interactions.

\section{The xenon recoil structure}
\label{sec:xenon}

\begin{figure}[!tbp]
\centering
\includegraphics[width=\columnwidth]{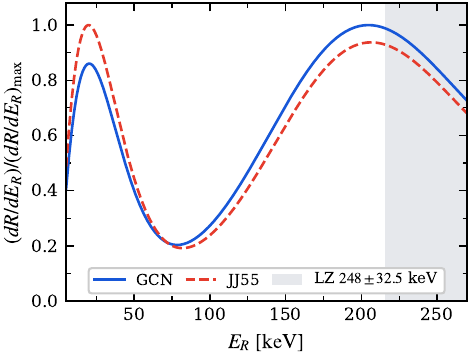}
\caption{Natural xenon recoil spectra for the correlated toroidal interaction at $m_X=1~\mathrm{TeV}$ in the contact regime. Each nuclear calculation is normalized to its largest value. The shaded interval is the quadrature combination of the reported statistical and systematic recoil uncertainties.}
\label{fig:spectrum}
\end{figure}

\Cref{fig:spectrum} shows the natural xenon recoil spectrum. GCN gives maxima at $20.3$ and $204.8~\mathrm{keV}$, whereas JJ55 gives $19.9$ and $206.6~\mathrm{keV}$. The LZ candidate lies $41$ to $43~\mathrm{keV}$ above the high recoil maximum, about $1.3$ times the quadrature energy uncertainty. Since this structure is broad, the rate retained at $248~\mathrm{keV}$ is more relevant: it is $85.4\%$ of the high maximum for GCN and $80.4\%$ for JJ55. The candidate therefore lies on the descending side of a broad high recoil branch rather than at a narrow endpoint.

The accompanying low recoil branch is smaller in the integrated response. We define
\begin{equation}
 \mathcal B(m_{A'})=
 \frac{\int_{100~\mathrm{keV}}^{250~\mathrm{keV}}dE_R\,R(E_R,m_{A'})}
 {\int_{14~\mathrm{keV}}^{100~\mathrm{keV}}dE_R\,R(E_R,m_{A'})}.
 \label{eq:branchratio}
\end{equation}
In the contact regime, \cref{eq:branchratio} gives $\mathcal B=3.10$ for GCN and $2.61$ for JJ55. Normalizing the high interval to unit response leaves only $0.322$ to $0.384$ response units in the low interval. One expected signal event in the high interval would therefore give a probability $e^{-1/\mathcal B}=0.724$ to $0.681$ for zero signal events in the low interval. We use this only as a response-level spectral consistency measure; an event likelihood requires detector-response folding.

The limits in \cref{eq:branchratio} also correspond closely to the published LZ response region. LZ reports an average nuclear recoil efficiency of $96\%$ between $14$ and $250~\mathrm{keV}$, with the efficiency falling below $50\%$ only outside $5.4$ to $269.9~\mathrm{keV}$ after all analysis criteria~\cite{LZ2026highE}. The collaboration likelihood is constructed in the detector variables $\{S1c,\log_{10}S2c\}$ from signal and background probability densities~\cite{LZ2026highE}. We therefore use $\mathcal B$ as a particle and nuclear response quantity over the high-efficiency interval. The pointwise ratio $H$ introduced below is likewise a true-recoil diagnostic and is not a substitute for the collaboration likelihood.

\begin{figure}[!tbp]
\centering
\includegraphics[width=\columnwidth]{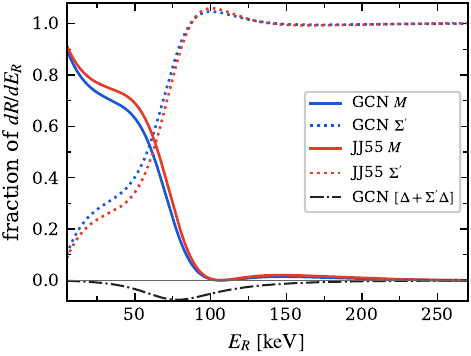}
\caption{Fractional nuclear response contributions for natural xenon. The coherent $M$ term dominates the first maximum. The $\Sp$ term dominates above $100~\mathrm{keV}$. The dotted curve shows the residual $\DD+\Sp\DD$ contribution for GCN.}
\label{fig:components}
\end{figure}

\Cref{fig:components} separates the responses forming the two maxima. At the GCN low recoil maximum, $M$ contributes $75.3\%$ and $\Sp$ contributes $25.3\%$; at the high maximum $\Sp$ contributes $99.7\%$. At $248~\mathrm{keV}$ the GCN fractions are $0.125\%$ from $M$, $0.00251\%$ from $\DD$, $99.93\%$ from $\Sp$, and $-0.0568\%$ from $\Sp\DD$. For JJ55 the corresponding $\Sp$ fraction is slightly above unity because the interference is negative. The high recoil branch is therefore governed by the transverse spin tensor response, with subpercent interference at the candidate energy.

The positive $M$ and $\Sp$ contributions are equal at $59.7~\mathrm{keV}$ for GCN and $63.6~\mathrm{keV}$ for JJ55. This crossover has only weak mediator dependence. As $m_{A'}$ is varied from $0.15~\mathrm{GeV}$ to the contact regime, it moves from $59.48$ to $59.68~\mathrm{keV}$ for GCN and from $63.45$ to $63.59~\mathrm{keV}$ for JJ55. The crossover energy therefore mainly probes the nuclear response, whereas the relative normalization of the two branches remains sensitive to the mediator.

\begin{figure}[!tbp]
\centering
\includegraphics[width=\columnwidth]{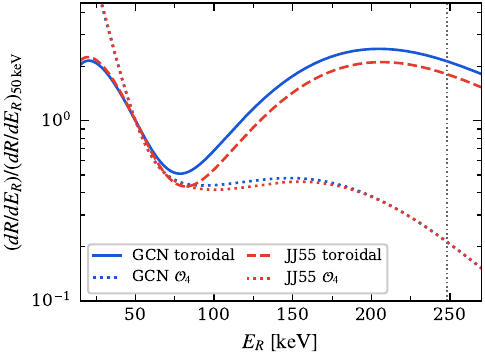}
\caption{Natural xenon spectra normalized at $50~\mathrm{keV}$. The toroidal interaction retains a high recoil branch, while the standard isovector $\mathcal O_4$ response decreases strongly through the candidate energy.}
\label{fig:o4}
\end{figure}

A conventional elastic spin interaction provides a useful reference. \Cref{fig:o4} compares the toroidal spectrum with the isovector $\mathcal O_4$ response using the same halo model and nuclear calculations. At $m_X=1~\mathrm{TeV}$,
\begin{equation}
 \begin{aligned}
 H_T&\equiv\frac{R_T(248)}{R_T(50)}=2.139\,(1.814),\\
 H_{\mathcal O_4}&=0.2131.
 \end{aligned}
 \label{eq:o4contrast}
\end{equation}
for GCN (JJ55). The conventional elastic spin response entering \cref{eq:o4contrast} has also been considered directly in interpretations of the current LZ data~\cite{ElahiSchwaller2026}. Relative to $\mathcal O_4$, the toroidal interaction increases the candidate-to-low-recoil ratio by a factor $10.0$ for GCN and $8.51$ for JJ55. This behavior follows from the $q^4$ tensor weighting together with the finite momentum $\Sp$ response.

\section{Mediator inference and analytical control}
\label{sec:boundary}

The spectral shape can be related directly to the mediator mass. Taking $E_H=248~\mathrm{keV}$ and $E_L=50~\mathrm{keV}$, we define
\begin{equation}
 H(m_X,m_{A'})=\frac{R(E_H)}{R(E_L)}.
 \label{eq:hardness}
\end{equation}
The propagator contribution to the ratio in \cref{eq:hardness} is
\begin{equation}
 H(m_{A'})\simeq H_\infty
 \left(\frac{m_{A'}^2+q_L^2}{m_{A'}^2+q_H^2}\right)^2,
 \label{eq:Hmediator}
\end{equation}
where $H_\infty$ denotes the full contact-response ratio. At $m_X=1~\mathrm{TeV}$, $H_\infty=2.139$ for GCN and $1.814$ for JJ55. The contact spectrum first reaches $H_\infty>1$ at $m_X=252~\mathrm{GeV}$ and $293~\mathrm{GeV}$, respectively.

The condition $H=1$ gives
\begin{equation}
 m_{A',\rm crit}^2=
 \frac{q_H^2-\sqrt{H_\infty}\,q_L^2}{\sqrt{H_\infty}-1}.
 \label{eq:mcrit}
\end{equation}
For $m_X=1~\mathrm{TeV}$, \cref{eq:mcrit} gives $m_{A',\rm crit}=0.302~\mathrm{GeV}$ for GCN and $0.355~\mathrm{GeV}$ for JJ55 in the natural isotope calculation. The representative isotope expression gives $0.304$ and $0.357~\mathrm{GeV}$, respectively. Over the heavy dark matter region the largest difference between the two evaluations is $0.54\%$.

The same relation also permits an inversion rather than only a lower bound. Define
\begin{equation}
 y=\sqrt{\frac{H}{H_\infty}}.
\end{equation}
Solving \cref{eq:Hmediator} for $m_{A'}$ yields
\begin{equation}
 \boxed{m_{A'}^2=\frac{yq_H^2-q_L^2}{1-y}}.
 \label{eq:invert}
\end{equation}
Its local sensitivity is
\begin{equation}
 \frac{\partial m_{A'}^2}{\partial y}=\frac{q_H^2-q_L^2}{(1-y)^2}.
 \label{eq:sensitivity}
\end{equation}
A measured spectral ratio can therefore determine $m_{A'}$ when the propagator is resolved. As $y\to1$, however, the contact limit is approached and the inferred mass becomes progressively less precise.

The lower bound is single valued because
\begin{equation}
 \frac{d\ln H}{d\ln m_{A'}}=4m_{A'}^2
 \left[\frac{1}{m_{A'}^2+q_L^2}-\frac{1}{m_{A'}^2+q_H^2}\right]>0
 \label{eq:monotonic}
\end{equation}
for $q_H>q_L$. The same reasoning applies to a more general explicit momentum dependence. For a spectrum
\begin{equation}
 R_n(E_R)=E_R^nG(E_R)
 \left(\frac{m_{A'}^2}{m_{A'}^2+q^2}\right)^2,
 \qquad q^2=\kappa E_R,
 \label{eq:generalrate}
\end{equation}
an interior extremum of \cref{eq:generalrate} satisfies
\begin{equation}
 -\frac{d\ln G}{d\ln E_R}
 =n-2+\frac{2m_{A'}^2}{m_{A'}^2+q^2}.
 \label{eq:generalpeak}
\end{equation}
For the tensor branch, $n=2$, and \cref{eq:generalpeak} reduces to
\begin{equation}
 -\frac{d\ln G}{d\ln E_R}=\frac{2m_{A'}^2}{m_{A'}^2+q^2}.
 \label{eq:peak}
\end{equation}
In the contact limit, the tensor condition in \cref{eq:peak} requires two units of logarithmic suppression from the nuclear response and halo phase space. In the light mediator limit the propagator compensates the explicit $q^4$ growth.

\begin{figure}[!tbp]
\centering
\includegraphics[width=\columnwidth]{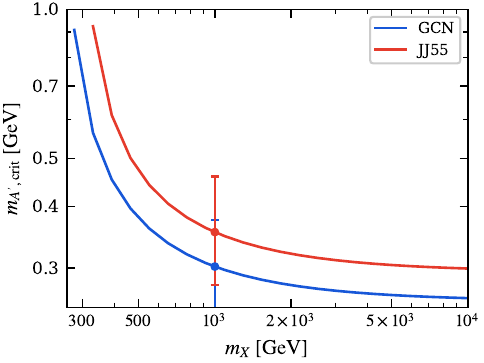}
\caption{Mediator mass required for $R(248)>R(50)$ as a function of $m_X$. The curves use the natural isotope calculation. The pale bands propagate the quadrature recoil energy interval through the pointwise boundary. The part above the LZ analysis endpoint should be interpreted only as recoil energy propagation.}
\label{fig:boundary}
\end{figure}

\Cref{fig:boundary} gives the mass dependence and the effect of the reconstructed recoil energy. At the central value, the nuclear interaction spread is $0.302$ to $0.355~\mathrm{GeV}$. Taking $E_H=215.5~\mathrm{keV}$ gives $0.240$ and $0.278~\mathrm{GeV}$, while the LZ endpoint at $270~\mathrm{keV}$ gives $0.375$ and $0.459~\mathrm{GeV}$. The upper quadrature value $280.5~\mathrm{keV}$ lies outside the reported analysis window and is included only as an extrapolative theory point. This energy dependence exceeds the standard halo variation discussed in \cref{sec:systematics}.

\begin{figure}[!tbp]
\centering
\includegraphics[width=\columnwidth]{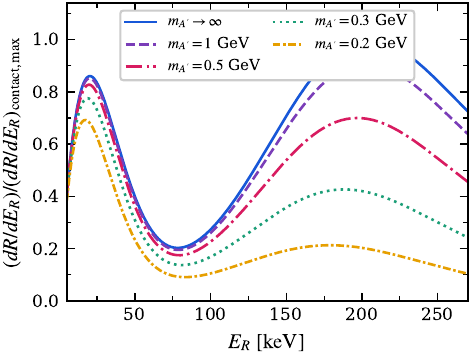}
\caption{GCN spectra for representative $m_{A'}$ at $m_X=1~\mathrm{TeV}$. Every curve uses the contact low recoil maximum as a common normalization. A lighter mediator suppresses the high momentum branch relative to the charge branch.}
\label{fig:finite}
\end{figure}

\begin{figure}[!tbp]
\centering
\includegraphics[width=\columnwidth]{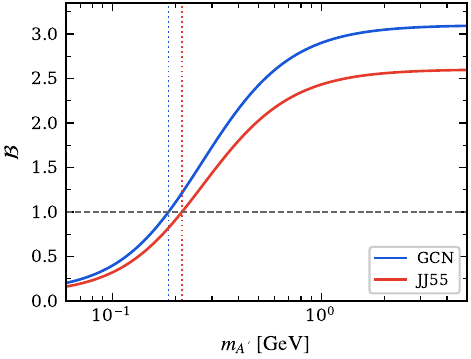}
\caption{Integrated branch ratio $\mathcal B$ at $m_X=1~\mathrm{TeV}$. The horizontal line marks equal high and low interval weights. The vertical dotted lines mark the stronger pointwise conditions from $H=1$.}
\label{fig:branch}
\end{figure}

\Cref{fig:finite} shows directly how the propagator modifies the spectrum. The integrated requirement $\mathcal B>1$ is weaker than the pointwise requirement $H>1$. It gives $m_{A'}>0.185~\mathrm{GeV}$ for GCN and $0.215~\mathrm{GeV}$ for JJ55, as displayed in \cref{fig:branch}.

\begin{figure}[!tbp]
\centering
\includegraphics[width=\columnwidth]{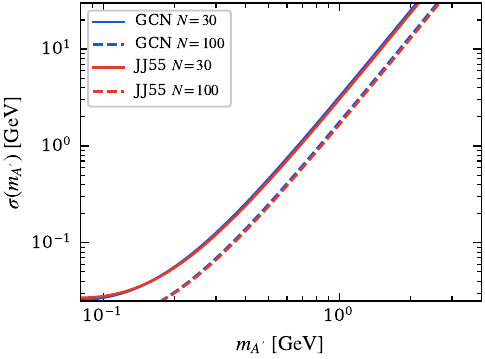}
\caption{Response level two bin uncertainty on $m_{A'}$ after profiling the total normalization. The mediator can be measured most efficiently near the propagator transition. The contact regime is primarily a lower bound problem.}
\label{fig:reach}
\end{figure}

With a larger event sample, the two recoil intervals can be used to estimate the mediator mass without fixing the absolute normalization. Let $p=\mathcal B/(1+\mathcal B)$ denote the high-interval fraction. For $N$ signal events, a two bin shape estimate gives
\begin{equation}
 \sigma(m_{A'})\simeq
 \frac{\sqrt{p(1-p)/N}}{|dp/dm_{A'}|}.
 \label{eq:fisher}
\end{equation}
At $m_{A'}=0.30~\mathrm{GeV}$ and $N=30$, the two bin estimate in \cref{eq:fisher} gives $\sigma(m_{A'})\simeq0.126~\mathrm{GeV}$ for GCN and $0.121~\mathrm{GeV}$ for JJ55. For $N=100$ the corresponding values are $0.069$ and $0.066~\mathrm{GeV}$. The uncertainty increases rapidly toward the contact limit, as expected from \cref{eq:sensitivity}. \Cref{fig:reach} shows this response-level information limit. A detector-level likelihood would additionally include acceptance, backgrounds, and nuisance parameters.

\section{Isotope and target tests}
\label{sec:isotopes}

The isotope content of the two branches follows from their different nuclear currents. The low branch contains $Q_p=1$ and $Q_n=0$, so its coherent component is present in every xenon isotope. The high branch depends on $g_p=5.59$ and $g_n=-3.83$ through the transverse spin response. Natural xenon has nonzero nuclear spin only in $^{129}\mathrm{Xe}$ and $^{131}\mathrm{Xe}$. Their finite momentum spin structure therefore dominates once the $q^4\Sp$ contribution exceeds the coherent term.

For GCN, the odd $A$ isotopes contribute $60.8\%$ at the low maximum, $99.70\%$ at the high maximum, and $99.94\%$ at $248~\mathrm{keV}$. JJ55 gives $58.9\%$, $99.62\%$, and $99.92\%$, respectively. At the candidate energy, $^{129}\mathrm{Xe}$ alone contributes $88.2\%$ for GCN and $89.7\%$ for JJ55. The high recoil region is therefore almost entirely an odd $A$ xenon response.

\begin{figure}[!tbp]
\centering
\includegraphics[width=\columnwidth]{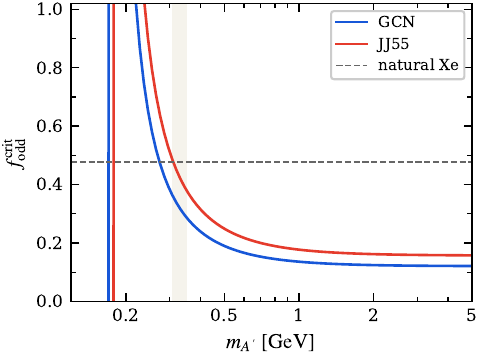}
\caption{Critical odd $A$ xenon fraction required for $R(248)>R(50)$ at $m_X=1~\mathrm{TeV}$. The shaded region above each curve satisfies the high recoil condition. The horizontal line gives natural xenon.}
\label{fig:isotopeboundary}
\end{figure}

Let $R_o$ and $R_e$ denote spectra for the odd and even isotope subsets, each normalized within its own subset. For a target with total odd fraction $f_{\rm odd}$,
\begin{equation}
 R(E)=f_{\rm odd}R_o(E)+(1-f_{\rm odd})R_e(E).
 \label{eq:mix}
\end{equation}
The linear combination in \cref{eq:mix} reaches the boundary $R(E_H)=R(E_L)$ at
\begin{equation}
 f_{\rm odd}^{\rm crit}=
 \frac{R_e(E_L)-R_e(E_H)}
 {[R_o(E_H)-R_o(E_L)]-[R_e(E_H)-R_e(E_L)]}.
 \label{eq:fcrit}
\end{equation}
At $m_X=1~\mathrm{TeV}$ in the contact regime, \cref{eq:fcrit} gives $f_{\rm odd}^{\rm crit}=0.166$ for GCN and $0.214$ for JJ55. Natural xenon has $f_{\rm odd}=0.477$, well above either value, as illustrated in \cref{fig:isotopeboundary}.

\begin{figure}[!tbp]
\centering
\includegraphics[width=\columnwidth]{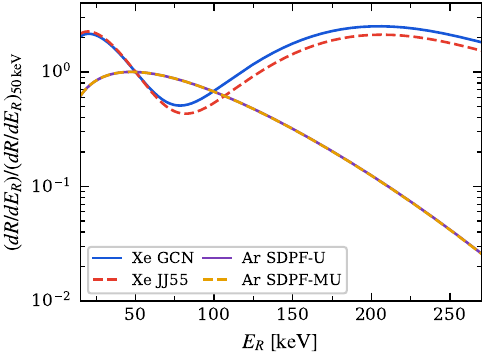}
\caption{Xenon and $^{40}\mathrm{Ar}$ contact spectra normalized at $50~\mathrm{keV}$. Xenon retains the tensor branch, while spin zero argon contains only the charge dominated branch.}
\label{fig:target}
\end{figure}

A spin zero target gives a direct test of the tensor contribution. The ground state of $^{40}\mathrm{Ar}$ has $J=0$, so the $\Otwo$ transverse spin tensor branch is absent, while the charge contribution from $\Osev$ remains. SDPF-U and SDPF-MU both give a peak near $47.4~\mathrm{keV}$ and
\begin{equation}
 H_{\rm Ar}=\frac{R_{\rm Ar}(248)}{R_{\rm Ar}(50)}=0.0438\text{ to }0.0439.
 \label{eq:Har}
\end{equation}
The small value of \cref{eq:Har} is a consequence of removing the tensor branch. The corresponding xenon-to-argon ratio is
\begin{equation}
 \mathcal D_{\rm Xe/Ar}=\frac{H_{\rm Xe}}{H_{\rm Ar}}
 \label{eq:double}
\end{equation}
in \cref{eq:double} reaches $41$ to $49$ in the contact regime. The full spectral comparison is shown in \Cref{fig:target}. Since the overall coupling cancels in this ratio while the halo dependence remains, we use it as a target-spin discriminator for the halo assumptions specified below.

\section{Timing and systematic hierarchy}
\label{sec:systematics}

\begin{figure}[!tbp]
\centering
\includegraphics[width=\columnwidth]{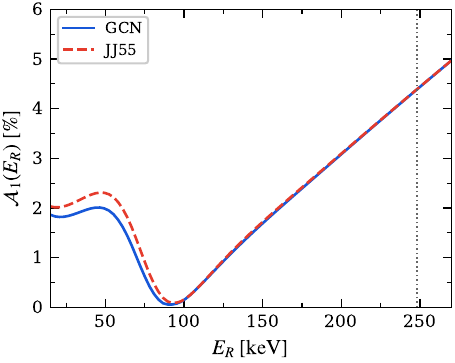}
\caption{Annual fractional modulation amplitude for $m_X=1~\mathrm{TeV}$ in the contact regime. The vertical line marks $248~\mathrm{keV}$.}
\label{fig:modulation}
\end{figure}

We adopt a truncated Maxwell distribution with $v_0=238~\mathrm{km\,s^{-1}}$ and $v_{\rm esc}=544~\mathrm{km\,s^{-1}}$. The annual modulation treatment follows the standard direct detection formalism~\cite{Drukier1986mod,Freese1988mod,Freese2013modulation}. Departures from a smooth Maxwellian halo and uncertainties in the local speed distribution can modify the modulation pattern~\cite{Savage2006streams,McCabe2010astro,Green2017astro}. Astrophysical nuisance parameters can be included in broader inference analyses~\cite{Frandsen2012astro}. We use the elastic kinematics and reporting conventions of Refs.~\cite{Lewin1995rx,Baxter2021pqo}. The annual fractional amplitude is
\begin{equation}
 A_1(E_R)=100\,\frac{R_{\max}(E_R)-R_{\min}(E_R)}{R_{\max}(E_R)+R_{\min}(E_R)}.
 \label{eq:moddef}
\end{equation}
Using \cref{eq:moddef}, we obtain $A_1=4.39\%$ at $248~\mathrm{keV}$ for both xenon calculations, as shown in \cref{fig:modulation}. Near the high recoil maximum the value is $3.2$ to $3.3\%$. At fixed recoil energy the mediator propagator is time independent and therefore cancels to very high accuracy in the fractional amplitude. Varying $m_{A'}$ from $0.2~\mathrm{GeV}$ to the contact regime changes $A_1(248)$ by less than $10^{-3}$ percentage point in our calculation. With the calendar convention of Ref.~\cite{Baxter2021pqo}, the rate at $248~\mathrm{keV}$ is maximal about $74.5$ days after March 22, approximately June 4, and this phase is unchanged over the mediator values considered above. The timing observable consequently probes the elastic halo kinematics almost independently of the mediator relation. Its few percent modulation differs from endothermic solutions close to the halo speed endpoint, where substantially larger seasonal fractions can arise~\cite{McCabe2026seasonal}.

The main systematics can be compared directly. Varying $v_0$ from $220$ to $250~\mathrm{km\,s^{-1}}$ and $v_{\rm esc}$ from $500$ to $560~\mathrm{km\,s^{-1}}$ changes the analytic pointwise mediator threshold from $0.300$ to $0.310~\mathrm{GeV}$ for GCN and from $0.352$ to $0.364~\mathrm{GeV}$ for JJ55. The integrated threshold changes by about $1~\mathrm{MeV}$ for GCN and $2~\mathrm{MeV}$ for JJ55. This halo variation is smaller than the GCN--JJ55 spread, while propagation of the candidate recoil energy gives a larger effect, as \cref{fig:boundary} shows.

The nuclear input used here is the one body electromagnetic current. Axial two body corrections in spin dependent WIMP studies correspond to a different current, so their uncertainty bands do not apply to the electromagnetic $\Sp$ response~\cite{Klos2013rwa,Hoferichter2015chiral}. Electromagnetic two body currents at $q\sim0.2$ to $0.25~\mathrm{GeV}$ remain a relevant extension. We therefore quote the GCN--JJ55 spread rather than importing an uncertainty band obtained for the axial current.

The uncertainty assigned to the candidate recoil energy is also different from the detector resolution. LZ quotes statistical and systematic recoil energy uncertainties and evaluates signal hypotheses in $\{S1c,\log_{10}S2c\}$ with nuisance-constrained background components~\cite{LZ2026highE}. The tested recoil spectra and calibrated detector response information are supplied in the public data release~\cite{LZ2026data}. Here we propagate the reported recoil energy interval only to determine how the theory-level pointwise relation changes with $E_H$. Extracting $g_T\epsilon$ from a single candidate requires a recast in detector observables or use of the collaboration likelihood. The response-level calculation therefore determines spectral ratios, while an absolute toroidal coupling can be inferred only after detector folding.

\subsection{High energy response tail}

The recoil range above the candidate region provides an additional probe of the propagator. We define the true-recoil response ratio
\begin{equation}
 \mathcal S_{\rm SB}(m_{A'})=
 \frac{\displaystyle\int_{350~\rm keV}^{590~\rm keV}dE_R\,R(E_R,m_{A'})}
 {\displaystyle\int_{100~\rm keV}^{270~\rm keV}dE_R\,R(E_R,m_{A'})}.
 \label{eq:sidebandratio}
\end{equation}
In the contact limit, $\mathcal S_{\rm SB}=0.740$ for GCN and $0.736$ for JJ55. At the gauge benchmark $m_{A'}=0.450~\mathrm{GeV}$, the propagator reduces these values to $0.463$ and $0.461$, while $m_{A'}=0.300~\mathrm{GeV}$ gives $0.337$ and $0.336$. The higher recoil response therefore carries mediator information beyond the $248~\mathrm{keV}$ point.

If the response in $100$ to $270~\mathrm{keV}$ is normalized to one expected event, the benchmark gives about $0.46$ response-level events in the $350$ to $590~\mathrm{keV}$ interval. The Poisson probability for zero events is then approximately $0.63$. Detector-level use of \cref{eq:sidebandratio} requires acceptance and observable-space folding. This matters because current sideband constraints on other LZ interpretations use experiment-specific information~\cite{Rodd2026sideband,DentNewstead2026}.

\section{Relation to current LZ interpretations}
\label{sec:comparisons}

Several mechanisms have now been considered for events in the vicinity of $250~\mathrm{keV}$. High scale supersymmetry and Peccei Quinn constructions give ultraviolet settings for electroweak dark matter~\cite{Yin2026lz,Visinelli2026lz}. Solar capture, neutrino telescopes, and the high energy sideband constrain Higgsino interpretations through independent observables~\cite{PospelovRamani2026lz,Bose2026nu,Rodd2026sideband}. Geometric multi target and ALP mediated scenarios introduce different target and momentum dependences~\cite{AhmedLeontaris2026,Yuan2026}. Exothermic and endothermic studies emphasize the information contained in the empty high energy sideband~\cite{DentNewstead2026}. A transition magnetic dipole gives a delayed photon accompanying the recoil~\cite{He2026transition}, while elastic neutrino scattering can be tested with both the event kinematics and the lower energy spectrum~\cite{Chattaraj2026nu}. The relevant distinguishing observables are summarized in \Cref{tab:compare}.

\begin{table}[!tbp]
\caption{Origin of the high recoil region and a discriminator for representative mechanisms.}
\label{tab:compare}
\scriptsize
\begin{tabular}{lll}
\hline\hline
Mechanism & Origin & Discriminator \\
\hline
Toroidal vector & $q^4\Sp$ response & \shortstack[l]{odd $A$ Xe, Ar branch loss,\\few percent timing} \\
$\mathcal L_{10}^{s}$ & momentum EFT & \shortstack[l]{double peak in the\\LZ search~\cite{LZ2026highE}} \\
$\mathcal O_4$ elastic & spin response & $H\simeq0.213$ at $1~\mathrm{TeV}$ \\
$\mathcal L_6^v$ elastic & nuclear interference & \shortstack[l]{Xe cancellation and\\target shift~\cite{Khan2026L6v}} \\
Endothermic & positive splitting & \shortstack[l]{threshold kinematics and\\halo tail~\cite{Su2026lz,Yamashita2026lz,Zhu2026lightDP}} \\
Exothermic & energy release & \shortstack[l]{sideband shape and\\argon response~\cite{BaerBarger2026,DentNewstead2026}} \\
Transition dipole & inelastic transition & \shortstack[l]{delayed recoil photon~\cite{He2026transition}} \\
Boosted & nonthermal flux & \shortstack[l]{incident spectrum and\\momentum response~\cite{Alhazmi2026,HeikinheimoZimmermann2026}} \\
Elastic $\nu$ & incident neutrinos & \shortstack[l]{kinematics plus low\\energy spectrum~\cite{Chattaraj2026nu}} \\
\hline\hline
\end{tabular}
\end{table}

The closest elastic alternatives differ in their nuclear response. The standard $\mathcal O_4$ interaction gives a much smaller high-to-low ratio in the same nuclear calculation. The elastic $\mathcal L_6^v$ interpretation instead produces a finite momentum cancellation near the event region and a target-dependent displacement of that cancellation~\cite{Khan2026L6v}. The toroidal interaction gives a broad high recoil maximum, a response crossover near $60~\mathrm{keV}$, and an almost pure odd $A$ xenon contribution at the candidate energy. These elastic cases are therefore distinguished by different nuclear mechanisms rather than by inelastic kinematics.

The comparison with inelastic dark photon models is also transparent. In those models the recoil scale is controlled primarily by a dark-state splitting close to a kinematic boundary~\cite{Yamashita2026lz,Zhu2026lightDP}. No such splitting appears here. Instead, the mediator determines how much of the tensor momentum dependence survives the propagator, while the position of the $M$--$\Sp$ crossover remains nearly fixed. The separation between a nuclear-response energy scale and a mediator-dependent branch ratio is therefore the characteristic test of the elastic toroidal scenario.

\section{Implications and conclusions}
\label{sec:conclusion}

The present LZ candidate is compatible, at the response level, with a broad elastic high recoil branch generated by the toroidal moment of spin $1$ dark matter. The GCN and JJ55 xenon calculations place the tensor maximum at $204.8$ to $206.6~\mathrm{keV}$ and retain $80.4$ to $85.4\%$ of this maximum at $248~\mathrm{keV}$. Over the $14$ to $250~\mathrm{keV}$ interval, where LZ reports a $96\%$ mean nuclear recoil efficiency, the integrated high recoil response is larger than the low recoil response by a factor $2.61$ to $3.10$. With one expected response-level event in the high interval, the corresponding low interval expectation is only $0.322$ to $0.384$ events. An isolated event in the extended window can therefore coexist with a smaller low recoil expectation without introducing a dark-state splitting.

The construction is constrained by electromagnetic matching, which fixes the relative $\Osev$ and $\Otwo$ coefficients. The $SU(2)_X\times U(1)_D$ realization relates vector stability, the dark photon mass, and the CP odd toroidal current while keeping the vector neutral under $U(1)_D$. The benchmark retains the high recoil tensor branch and gives $\mathcal S_{\rm SB}\simeq0.46$. At the event energy more than $99.9\%$ of the response is odd $A$ xenon, whereas spin zero argon lacks the tensor branch. The standard $\mathcal O_4$ interaction gives a candidate-to-low-recoil ratio almost an order of magnitude smaller. Equation~\eqref{eq:zetabound} also constrains softer coherent interactions that dilute the hard branch.

The mediator dependence admits a simple analytic description. At $m_X=1~\mathrm{TeV}$, the central recoil energy gives $m_{A'}>0.302$ to $0.355~\mathrm{GeV}$ for $R(248)>R(50)$. When future data resolve the propagator transition, Eq.~\eqref{eq:invert} converts this lower limit into an estimator of $m_{A'}$. Equation~\eqref{eq:generalpeak} gives the analogous extremum condition for a general explicit momentum power. Standard halo variations shift the mediator threshold by only about $0.01~\mathrm{GeV}$ within each xenon calculation, while the reconstructed candidate energy and the nuclear interaction choice produce larger variations.

A larger high recoil sample can test the different ingredients separately. The xenon spectrum is expected to contain a coherent-to-tensor crossover near $60$ to $64~\mathrm{keV}$ together with a broad high recoil branch dominated by $^{129}\mathrm{Xe}$ and $^{131}\mathrm{Xe}$. For a spin zero argon target the tensor branch is absent. The elastic halo model gives a modulation at the few percent level, while the population above $350~\mathrm{keV}$ probes the mediator through \cref{eq:sidebandratio}. Once events populate separated recoil intervals, ratios of their populations can constrain the dark photon mass without first determining the absolute normalization. These observables separate the nuclear, mediator, and halo dependences of the model.

\section*{Data availability}
The numerical implementation, response tables, and parameter scans used for the figures are supplied with the ancillary material. The xenon one body density matrices are publicly available through \texttt{dmscatter}~\cite{Gorton2022cpc}, and the experimental release associated with the extended LZ search is archived in HEPData~\cite{LZ2026data}.

\section*{Acknowledgements}
  SC  acknowledges the support of  Istituto Nazionale di Fisica Nucleare (INFN) Sezioni  di Napoli e di Frascati, {\it Iniziative Specifiche} QGSKY,  MOONLIGHT2 and  the Gruppo Nazionale di Fisica Matematica (GNFM)  of Istituto Nazionale di Alta Matematica (INDAM). 

\appendix
\section{Matching and response benchmarks}
\label{app:checks}

The electromagnetic matching may be displayed before performing the nuclear contraction. For an on shell nucleon, we write the current as
\begin{equation}
\begin{aligned}
 \langle N(p')|J_{\rm EM}^\mu|N(p)\rangle
 &=\bar u(p')\Bigg[F_1^N(q^2)\gamma^\mu \\
 &\quad+\frac{iF_2^N(q^2)}{2m_N}\sigma^{\mu\nu}q_\nu\Bigg]u(p),
\end{aligned}
 \label{eq:emcurrent}
\end{equation}
with $F_1^N(0)=Q_N$ and $2[F_1^N(0)+F_2^N(0)]=g_N$. Using the electromagnetic equation of motion in \cref{eq:oxg}, expanding \cref{eq:emcurrent} in the nonrelativistic limit, and projecting the vector polarizations onto \cref{eq:tensor}, one obtains
\begin{equation}
 \OXg\longrightarrow 2 e m_N^2\left(2Q_N\Osev+g_N\Otwo\right),
 \label{eq:matchingappendix}
\end{equation}
which reproduces \cref{eq:matching}. The charge term is governed by $F_1^N(0)$, whereas the transverse magnetic term contains the complete static magnetic moment $F_1^N(0)+F_2^N(0)$. Inserting $c_{17}^N=2Q_N$ and $c_{20}^N=g_N$ into the spin $1$ dark matter response functions gives \cref{eq:rm,eq:rd,eq:rs,eq:rsd} and hence the recoil expression in \cref{eq:rate}.

Before carrying out the velocity average, the four contributions can be combined into the convention check
\begin{equation}
\begin{aligned}
 \mathcal R_A&\propto\frac{x^2}{6}\sum_{NN'}\Bigg[
 4Q_NQ_{N'}W_{\Delta}^{NN'}
 +4Q_NQ_{N'}\frac{v_T^{\perp2}}{x}W_M^{NN'}\\
 &\hspace{2.6cm}+\frac14 g_Ng_{N'}W_{\Sigma'}^{NN'}
 -g_NQ_{N'}W_{\Sigma'\Delta}^{NN'}\Bigg].
\end{aligned}
 \label{eq:amplitudecheck}
\end{equation}
Equation~\eqref{eq:amplitudecheck} is the response decomposition of the published relativistic-to-nonrelativistic matching~\cite{Liang2025vector}. It displays explicitly the origin of the $M$, $\Delta$, $\Sigma'$, and $\Sigma'\Delta$ channels. The first two originate from $\mathcal O_{17}$, the third from $\mathcal O_{20}$, and the last from their interference. In this form the apparently additional $\Delta$ term is seen to be part of the complete $\mathcal O_{17}$ dark matter response.

For reproducibility, \cref{tab:bench} gives the contracted nuclear functions entering \cref{eq:rate} at representative momenta. We define $\widehat W_K$ to include the factor $(2J_A+1)^{-1}$ and the contractions with $c_{17}=(2,0)$ and $c_{20}=(g_p,g_n)$; isotope abundances and halo moments are included separately. The even $A$ row is averaged using the natural relative abundances within the spin zero subset. The complete numerical table is supplied with the accompanying data.

\begin{table*}[!tbp]
\caption{Contracted response benchmarks. Momentum is in MeV. All entries use the convention defined above.}
\label{tab:bench}
\centering
\tiny
\begin{tabular}{llrrrr}
\hline\hline
Interaction and target & $q$ & $\widehat W_M$ & $\widehat W_\Delta$ & $\widehat W_{\Sigma'}$ & $\widehat W_{\Sigma'\Delta}$ \\
\hline
\multicolumn{6}{c}{GCN} \\
$^{129}$Xe & 50 & 573 & $4.28\times10^{-4}$ & 0.615 & 0.0162 \\
$^{131}$Xe & 50 & 572 & $8.80\times10^{-4}$ & 0.135 & -0.0109 \\
even $A$ Xe & 50 & 571 & 0 & 0 & 0 \\
$^{129}$Xe & 100 & 113 & $1.32\times10^{-4}$ & 0.0807 & $3.27\times10^{-3}$ \\
$^{131}$Xe & 100 & 112 & $2.69\times10^{-4}$ & $3.25\times10^{-4}$ & $-1.71\times10^{-4}$ \\
even $A$ Xe & 100 & 111 & 0 & 0 & 0 \\
$^{129}$Xe & 150 & 1.46 & $1.45\times10^{-5}$ & 0.0114 & $4.07\times10^{-4}$ \\
$^{131}$Xe & 150 & 1.33 & $2.96\times10^{-5}$ & 0.0170 & $5.46\times10^{-4}$ \\
even $A$ Xe & 150 & 1.21 & 0 & 0 & 0 \\
$^{129}$Xe & 200 & 1.39 & $1.79\times10^{-7}$ & 0.0243 & $6.59\times10^{-5}$ \\
$^{131}$Xe & 200 & 1.43 & $1.01\times10^{-6}$ & 0.0168 & $1.14\times10^{-4}$ \\
even $A$ Xe & 200 & 1.48 & 0 & 0 & 0 \\
$^{129}$Xe & 250 & 0.103 & $5.83\times10^{-9}$ & 0.0207 & $-1.10\times10^{-5}$ \\
$^{131}$Xe & 250 & 0.0999 & $1.58\times10^{-7}$ & $2.98\times10^{-3}$ & $1.95\times10^{-5}$ \\
even $A$ Xe & 250 & 0.0975 & 0 & 0 & 0 \\
\hline
\multicolumn{6}{c}{JJ55} \\
$^{129}$Xe & 50 & 573 & $3.98\times10^{-4}$ & 0.524 & 0.0144 \\
$^{131}$Xe & 50 & 572 & $1.13\times10^{-3}$ & 0.101 & -0.0107 \\
even $A$ Xe & 50 & 571 & 0 & 0 & 0 \\
$^{129}$Xe & 100 & 113 & $1.23\times10^{-4}$ & 0.0638 & $2.80\times10^{-3}$ \\
$^{131}$Xe & 100 & 112 & $3.44\times10^{-4}$ & $9.42\times10^{-5}$ & $1.37\times10^{-4}$ \\
even $A$ Xe & 100 & 111 & 0 & 0 & 0 \\
$^{129}$Xe & 150 & 1.42 & $1.35\times10^{-5}$ & $7.55\times10^{-3}$ & $3.20\times10^{-4}$ \\
$^{131}$Xe & 150 & 1.27 & $3.71\times10^{-5}$ & 0.0136 & $5.87\times10^{-4}$ \\
even $A$ Xe & 150 & 1.17 & 0 & 0 & 0 \\
$^{129}$Xe & 200 & 1.43 & $2.21\times10^{-7}$ & 0.0186 & $6.41\times10^{-5}$ \\
$^{131}$Xe & 200 & 1.50 & $1.31\times10^{-6}$ & 0.0122 & $1.14\times10^{-4}$ \\
even $A$ Xe & 200 & 1.52 & 0 & 0 & 0 \\
$^{129}$Xe & 250 & 0.107 & $1.92\times10^{-12}$ & 0.0163 & $1.77\times10^{-7}$ \\
$^{131}$Xe & 250 & 0.106 & $2.68\times10^{-7}$ & $1.98\times10^{-3}$ & $2.02\times10^{-5}$ \\
even $A$ Xe & 250 & 0.101 & 0 & 0 & 0 \\
\hline\hline
\end{tabular}
\end{table*}

The mediator relation has two limiting forms. For $m_{A'}\gg q_H$, $H\to H_\infty$, whereas for $m_{A'}\ll q_L$,
\begin{equation}
 H\to H_\infty\left(\frac{q_L}{q_H}\right)^4,
 \label{eq:lightlimit}
\end{equation}
so that the propagator compensates the explicit $q^4$ enhancement of the tensor branch. The positivity of \cref{eq:monotonic} implies a single crossing of $H=1$ whenever the contact result has $H_\infty>1$ and the light-mediator limit in \cref{eq:lightlimit} lies below unity.

For the branch ratio, independent interpolation with adaptive quadrature gives $\mathcal B=3.10087$ for GCN and $2.60426$ for JJ55, compared with $3.1020$ and $2.6050$ obtained by direct integration on the recoil grid. The relative difference is below $0.04\%$. The analytic mean inverse speed likewise agrees with direct velocity quadrature over the recoil and mass range used in the analysis.

\setlength{\bibsep}{-0.15pt}
\bibliography{references}
\end{document}